\documentclass[prb,amsmath,amssymb]{revtex4}

\usepackage{lipsum}
\usepackage{booktabs}
\usepackage{graphicx}% Include figure files
\usepackage{dcolumn}% Align table columns on decimal point
\usepackage{bm}% bold math
\usepackage{lmodern}
\usepackage{siunitx}

\begin{document}

\title{Domain wall magneto-Seebeck effect}

\author{Patryk Krzysteczko}
\email{patryk.krzysteczko@ptb.de.de}
\author{Xiukun Hu}
\author{Niklas Liebing}
\author{Sibylle Sievers}
\author{Hans W. Schumacher}
\affiliation{Physikalisch-Technische Bundesanstalt, Bundesallee 100, D-38116 Braunschweig, Germany}

\begin{abstract}
The interplay between charge, spin, and heat currents in magnetic nano systems subjected to a temperature gradient has lead to a variety of novel effects and promising applications studied in the fast-growing field of spincaloritronics. Here we explore the magnetothermoelectrical properties of an individual magnetic domain wall in a permalloy nanowire. In thermal gradients of the order of few \si{K\per\micro m} along the long wire axis, we find a clear magneto-Seebeck signature due to the presence of a single domain wall. The observed domain wall magneto-Seebeck effect can be explained by the magnetization-dependent Seebeck coefficient of permalloy in combination with the local spin configuration of the domain wall.
\end{abstract}

\maketitle

Electronic transport coefficients in ferromagnetic materials are spin-dependent \cite{Mott1953book} enabling important spintronics applications \cite{Zutic2004rmp}. This observation also holds for magnetothermoelectric (or spincaloritronic) phenomena \cite{Johnson1987prb,Bauer2010ssc,Bauer2012nmat}, driven by thermal gradients \cite{Uchida2008nature,Slachter2010nphys,LeBreton2011nature,Jeon2014nmat}. In a thermal gradient, the temperature difference $\Delta T$ between two contacts gives rise to a thermopower $V_\mathrm{T}=-S\Delta T$ with $S$ being the material's Seebeck coefficient. Spin-dependent Seebeck coefficients have been observed in various nanomagnetic systems like thin films \cite{Avery2012prl,Schmid2013prl}, multilayers \cite{Shi1993jap}, tunnel junctions \cite{Czerner2011prb,Walter2011nmat,Liebing2011prl}, and nanowires \cite{Bohnert2013apl,Gravier2006prb}. In the latter, magnetization reversal often occurs by the nucleation and propagation of a single magnetic domain wall (DW) enabling promising applications \cite{Ono1999science,Allwood2005science,Parkin2008science}. Also a DW can interact with a thermal gradient \cite{Berger1985apl,Hatami2007prl,Kovalev2009prb} with prospects for thermally driven DW motion \cite{Hinzke2011prl,Yan2011prl,Torrejon2013prl,Jiang2013prl} or nanoscale magnetic heat engines \cite{Bauer2010prb}. However, the fundamental thermoelectrical properties of an individual magnetic DW have not been investigated yet.

\begin{figure}
 	\includegraphics[scale=1]{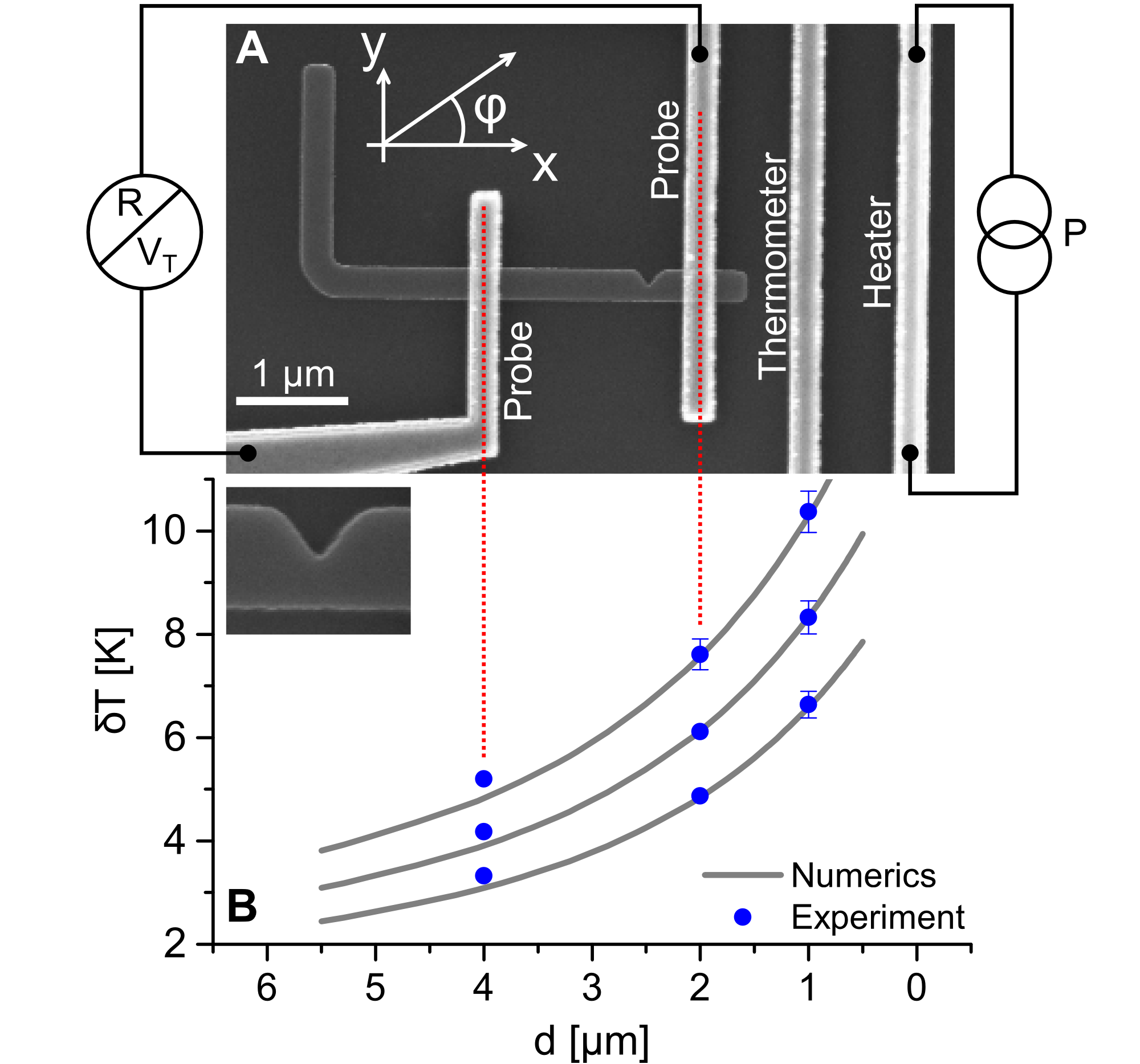}
	\caption{Sample geometry and temperature distribution. (A) Micrograph of the L-shaped permalloy nanostructure with Pt contact probes for voltage and resistance measurement. The inset below shows the notch with higher magnification. (B) The temperature increase for three heater powers $P=17$, $22$, and $27\si{mW}$ as a function of the distance to the heater. Experimental results are shown by blue bullets; numerical results (grey lines) show a good agreement.} 
	\label{i}
\end{figure}

\begin{center}
\begin{figure}
 	\includegraphics[scale=1]{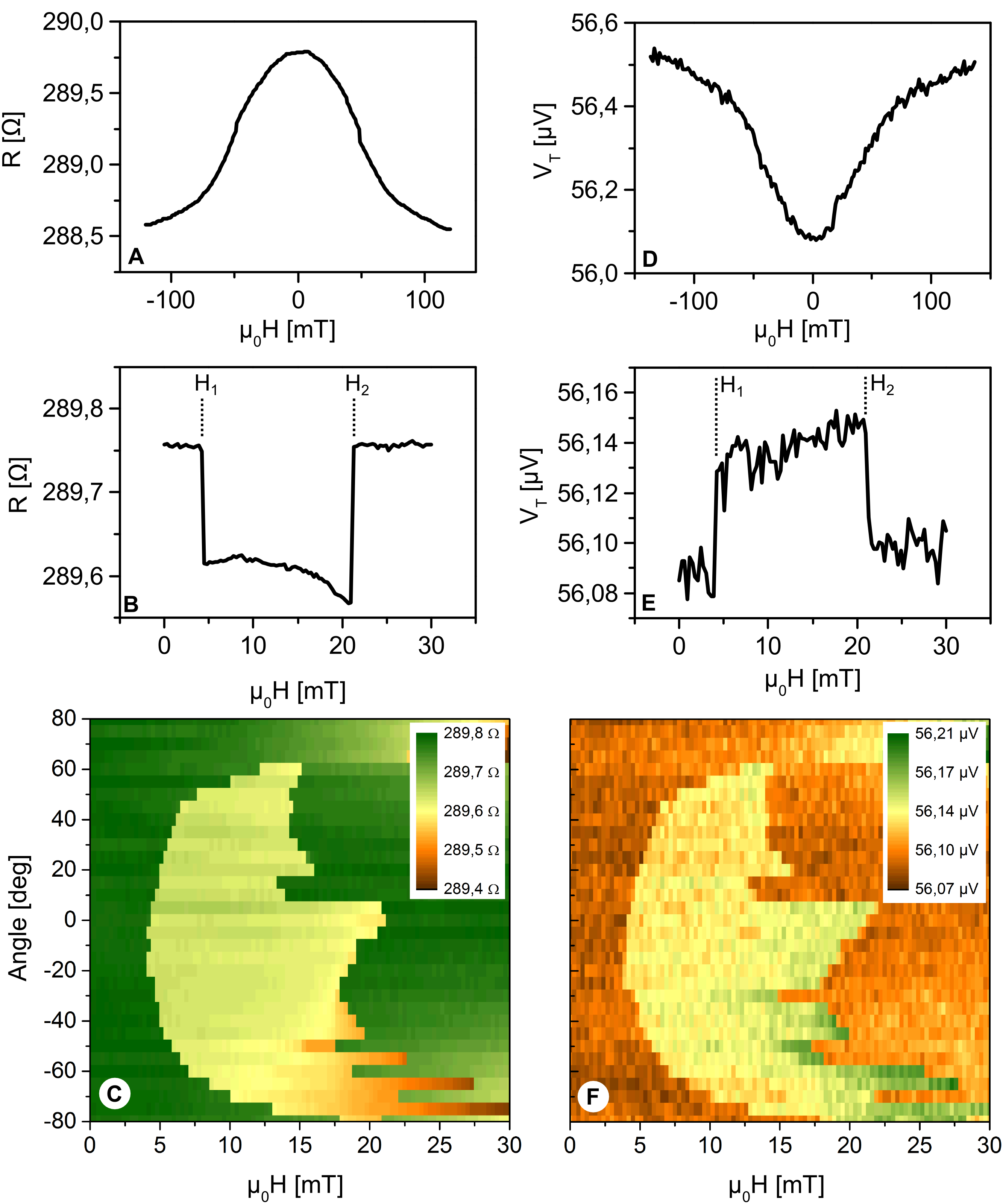} 
	\caption{Magnetoresistance and magnetothermopower measurement. (A) The resistance $R$ of the wire vs.\ applied field $\mu_0H$ measured in transversal geometry ($\phi=\ang{90}$) (B) Domain wall magnetoresistance measured at $\phi=\ang{0}$. The pinning fields of the corner and the notch are indicated by $H_1$ and $H_2$, respectively. (C) A set of domain wall magnetoresistance measurements for angles $|\phi|<\ang{80}$. The resistance is indicated by the color scale. (D) The thermopower of the wire measured in transversal geometry ($\phi=\ang{90}$) (E) Domain wall magnetoresistance measured at $\phi=\ang{0}$. (F) A set of domain wall magnetothermopower measurements for all angles $|\phi|<\ang{80}$. The thermopower is indicated by the color scale.} 
	\label{ii}
\end{figure}
\end{center}

\begin{figure}
 	\includegraphics[scale=1.2]{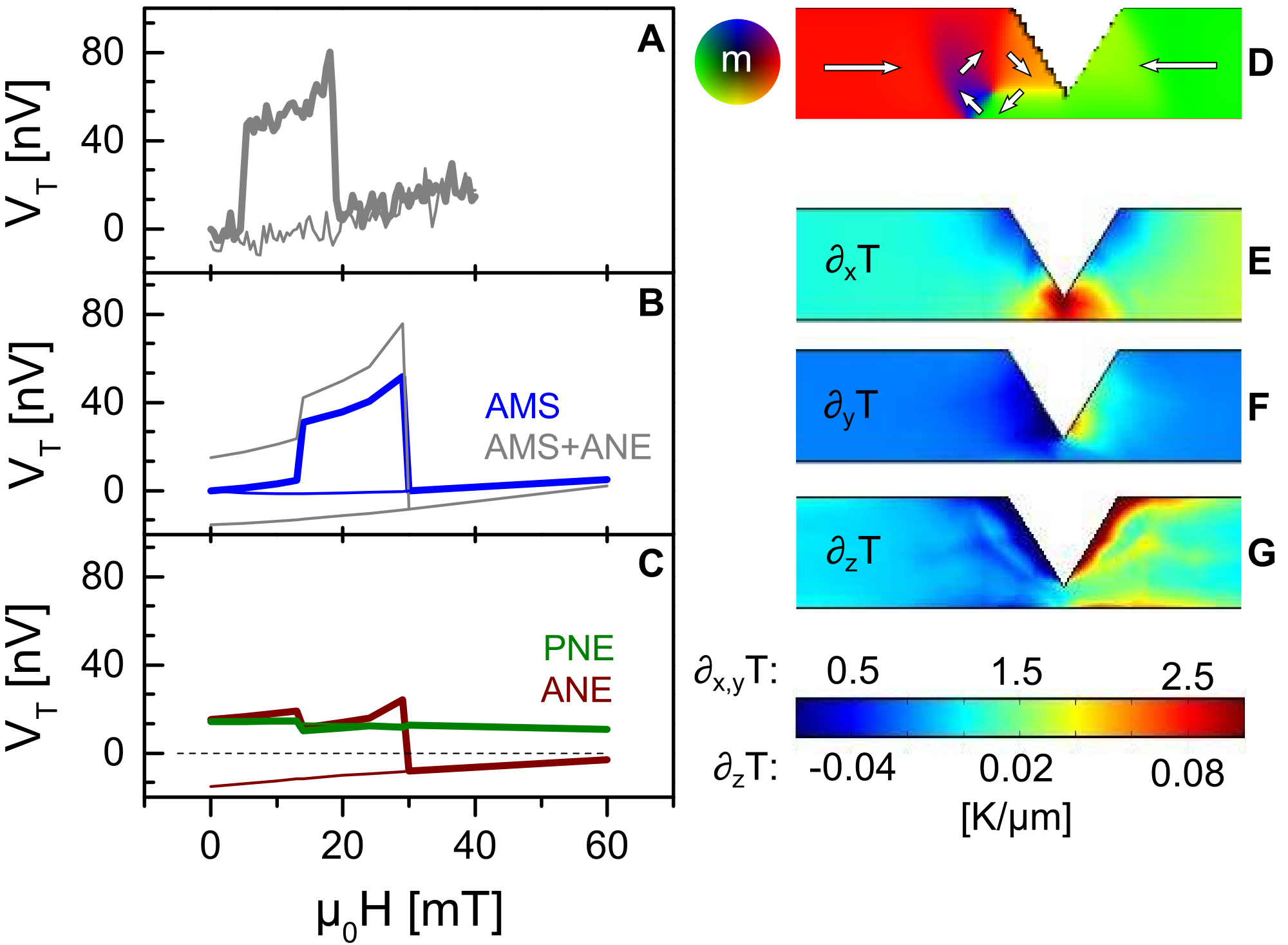}
	\caption{(A) Domain wall magnetothermopower measured at $\phi=\ang{-30}$. The graph shows the thermopower change with respect to the remanent state $V_{\mathrm{T}\parallel}$. (B) The calculated AMS contribution (blue) and the calculated contribution due to AMS and ANE acting together (grey). (C) Calculated contribution due to PNE (green) and ANE (brown). (D) Example of a simulated magnetization distribution showing a vortex DW. The local magnetization direction is indicated by the color scale. (E-G) The calculated temperature gradient in $x$, $y$ and $z$-direction.} 
	\label{iii}
\end{figure}

In our experiments we use L-shaped permalloy (Py) nanowires with a notch (see Fig.~\ref{i}A and supplementary material for details). The L's corner allows a controlled nucleation of a DW while the notch allows pinning a moving DW between the electrical probes. The two probes are contacting the Py wire from the top for resistance and thermopower measurements. Two additional Pt strips located at a distance of $0.5\,\si{\micro m}$ and $1.5\,\si{\micro m}$ from the Py nanowire serve as resistive thermometer and heater, respectively. The magnetic behavior of the system is characterized by two-wire resistance measurements as a function of magnetic field at a DC current of $600\,\si{\micro A}$. In a first step, the magnetization of the entire wire is rotated from the longitudinal ($\parallel$) to the transversal ($\perp$) direction by a magnetic field applied at $\phi=\ang{90}$, i.e. along the $y$-direction (note the definition of coordinates in Fig.~\ref{i}A). As expected for a system dominated by the anisotropic magnetoresistance (AMR), the measurement shows a bell-shaped curve (Fig.~\ref{ii}A) with resistance being decreased by the field of either polarity by $\Delta R=R_\parallel-R_\perp$. We find $R_\parallel=289.8\,\si{\ohm}$ at remanence and $R_\perp=288.5\,\si{\ohm}$ at maximum transversal field and hence a two-wire AMR ratio $\Delta R/R_\perp=0.45\,\%$. In a second step, we study the AMR contribution of a single DW. For this purpose we apply a $120\,\mathrm{mT}$ field in diagonal direction ($\phi_\mathrm{set}=\ang{-135}$) to create a head-to-head DW at the corner which is than moved towards the notch by a field applied at any $|\phi|<\ang{80}$. As an example, Fig.~\ref{ii}B shows a measurement at $\phi=\ang{0}$. The DW arrives at the notch at $H_1$, where it remains until $H_2$ is reached. The presence of the DW at the notch leads to a decrease of resistance by approximately $0.17\,\si{\ohm}$. The resistance drop is due to transversally oriented magnetization within the DW and based entirely on AMR. The critical fields $H_1$ and $H_2$ are the pinning fields of the corner and of the notch, respectively. To fully characterize the DW dynamics we repeat the measurement in the angle-range $|\phi|<\ang{80}$. The results are presented in Fig.~\ref{i}C, where the resistance is indicated by a color-scale. The yellow region indicates the resistance lowered due to the presence of the DW at the notch. Typically the left edge of this region is smooth whereas the right edge is rather irregular. This means that the pinning strength of the corner $H_1(\phi)$ for various angles is well defined \cite{Corte2014srep} whereas the pinning strength of the notch $H_2(\phi)$ has a stronger stochastic component. We model the magnetization distribution during field-driven DW motion by micromagnetic simulations using a Landau-Lifshitz-Gilbert micromagnetic simulator \cite{Scheinfein1991prb}. Our numerical analysis predicts that a vortex-type of DW is nucleated at the corner as pictured in Fig.~\ref{iii}D, where a snapshot of the magnetization distribution at $\mu_0H=20\,\si{mT}$ during a field sweep at $\phi=\ang{-30}$ is shown. For increasing field strength, the vortex DW will be 'pulled' deeper into the notch, deformed and finally transformed into a transversal DW before depinning, which explains the stochastic behavior of $H_2(\phi)$.

%1
%\onecolumngrid
%\begin{center}

For thermoelectrical measurements, we generate temperature gradients by applying an AC power $P$ at a frequency of $f=262\,\si{Hz}$ to the heater. To characterize the temperature distribution we use calibration samples with identical heaters and thermometers placed correspondingly to the positions of the voltage probes (red lines in Fig.~\ref{i}A). For each heater power $P$, the thermometer resistance has a $2f$ AC component with amplitude $\delta R(P)$ detected by 4-wire lock-in measurements. To translate $\delta R$ to the temperature increase $\delta T$ we first determine the temperature coefficient $\alpha_\mathrm{Pt}$ in a separate setup. We find $\alpha_\mathrm{Pt}=0.0013\,\si{\per\kelvin}$ which is 30\,\% of the bulk value in good agreement with literature \cite{Zhang2005apl}. Figure~\ref{i}B shows the measured $\delta T$ (blue bullets) as a function of the distance $d$ from the heater for three heating powers: $17\,\si{mW}$, $22\,\si{mW}$ and $27\,\si{mW}$. The temperature distribution is further investigated by three-dimensional finite-element modeling. The numerical results (gray lines) show a good agreement with the experimental data. Heating with $27\,\si{mW}$ accordingly leads to an increase of the nanowire temperature of up to $10\,\si{K}$ and a $\Delta T$ between the probes of $(2.4\pm0.5)\,\si{K}$. In the following, the thermopower $V_\mathrm{T}$ is measured at $P=27\,\si{mW}$ by lock-in detection at $2f$ via the voltage probes.

%2

Figure~\ref{ii}D shows the evolution of the thermopower $V_\mathrm{T}$ as a function of transversal field ($\phi=\ang{90}$, cf.\ Fig.~\ref{ii}A). Again, a bell-shaped curve comes into view, but with the thermopower being \emph{increased} by a magnetic field of either polarity. We find a thermopower of $V_\mathrm{T\parallel}=56.08\,\si{\micro V}$ at remanence and $V_\mathrm{T\perp}=56.54\,\si{\micro V}$ at maximum field with an accuracy of $\pm 10\,\si{nV}$. The effective Seebeck coefficient is $S=(23\pm 6)\,\si{\micro V/K}$. The magnetothermopower (MTP) ratio $(V_\mathrm{T\parallel}-V_\mathrm{T\perp})/V_\mathrm{T\perp}$ yields $(-0.81\pm 0.03)\,\%$. The Seebeck coefficient of the nanowire thus rises when the wire's magnetization rotates under the action of an external field. For comparison of magnetoresistance (MR) and magento-Seebeck ratio the lead contributions have to be taken into account as discussed in the supplementary material. In the following we investigate the change of thermopower induced by the presence of a single DW. As an example, Fig.~\ref{ii}E shows a MTP measurement at the same conditions as the MR measurement shown in Fig.~\ref{ii}B. As the field reaches $\mu_0H=4\,\si{mT}$, we observe a sudden increase of thermopower by approx.\ $40\,\si{nV}$. The thermopower remains roughly constant at this level until the field reaches $21\,\si{mT}$, where it drops back to the base level. Figure~\ref{ii}F shows the complete set of DW thermopower (DWTP) measurements for angles $|\phi|<\ang{80}$. In this color plot, the yellow area indicates the increased thermopower. If we compare the pinning fields from MR and thermopower measurements (Figs.~\ref{ii}C and \ref{ii}F), and keep the stochastic nature of $H_2$ in mind, we can safely consider them as identical. Evidently the origin of increased thermopower is the same as the origin of reduced resistance, namely the presence of a DW at the notch. The data thus clearly reveal the thermoelectrical signature of a single DW. 

To analyze our data, we describe the thermopower of a system magnetized along the $x$-direction by
\begin{equation}
\nabla V_\mathrm{T}=-
\begin{pmatrix}
S_\parallel &  & \\
 & S_\perp & -S_N\\
 & S_N & S_\perp
\end{pmatrix}
\nabla T\,,
\label{eq_matrix}
\end{equation}
where the Seebeck coefficient has tensor character analogous to the resistivity tensor (see supplementary material). The diagonal elements of the tensor represent the anisotropy of the Seebeck coefficient; $S_\parallel$ is measured when the temperature gradient is parallel to the magnetization direction while $S_\perp$ is measured when it is transversal to the magnetization direction (cf.\ Fig.~\ref{ii}d). We consider also the anomalous Nernst effect (ANE) by the off-diagonal elements $S_N=-2.6\,\si{\micro V\per K}$, which will generate an additional thermopower in the case of a non-vanishing out-of-plane temperature gradient \cite{Slachter2011prb}. Our experimental setup is designed to detect the thermopower generated along the wire direction thus we consider only the $x$-component of Eq.~\ref{eq_matrix}. The resulting MTP can be described by three terms
\begin{eqnarray}
 V_\mathrm{T} & = & -\left(S_\perp+\Delta S\cos^2(\theta)\right)\Delta T_x \nonumber \\
                        &  & -\Delta S\cos(\theta)\sin(\theta)\,\Delta T_y \nonumber \\
                        &  & -S_N\sin(\theta)\,\Delta T_z\,,
\label{eq_mtp}                        
\end{eqnarray}
where $\theta$ is the angle of the local magnetization direction with respect to the $x$-direction (see supplementary material). Due to the analogy with AMR, we refer to the first term as anisotropic magneto-Seebeck (AMS) effect. The second term is related to the planar Nernst effect (PNE) \cite{Avery2012prl} and the third term describes the ANE contribution of an in-plane magnetized system. We use our numerical results of magnetization distribution and temperature gradients to verify this approach. The nanowire is divided in cells of $10\times10\,\si{nm^2}$. For each cell we take the local magnetization direction $\theta(x,y)$ and the temperature difference across the cell to calculate the local thermopower according to Eq.~\ref{eq_mtp}. To estimate the global thermopower, we calculate the mean thermopower generated in each 10-nm-slice of the wire and sum over all slices between the voltage probes. We repeat those calculations for various magnetic configurations (cf.\ Fig.~\ref{iii}D) corresponding to the movement of a DW during a field-sweep at $\phi=\ang{-30}$. The temperature gradients are shown as a color map in Fig.~\ref{iii}E--G (note the different color scales for in-plane and out-of-plane directions). 

%3

Figure~\ref{iii}A shows the measured DWTP for $\phi=\ang{-30}$ with the thermopower of the remanent state $V_\mathrm{T\parallel}$ set to zero. The calculation considering only the AMS is shown in Fig.~\ref{iii}B by the blue curve. Considering the typical deviations of a micromagnetic model a very good agreement between experiment and simulation is obtained. Our analysis thus reveals that the DWTP is dominated by the AMS (first term in Eq.~\ref{eq_mtp}) and the remaining terms of Eq.~\ref{eq_mtp} are treated as corrections. The expected PNE contribution is indicated by the green line in Fig.~\ref{iii}C. It shows a nearly constant value of approx.\ $15\,\si{nV}$ with a DW signature of only $3\,\si{nV}$. Within the experimental noise level, PNE should hence have no impact on the DWTP. Figure~\ref{iii}C also plots the ANE contribution (brown line). At $\mu_0H=0\,\si{mT}$ the ANE has a value of about $15\,\si{nV}$. During the DW pinning only a small change of the ANE signal occurs. However, at $30\,\si{mT}$ the ANE signal shows a sudden drop and a change of sign. Here the magnetization direction at the hot side of the notch is reversed due to the depinning of the DW. In the following up-sweep to $60\,\si{mT}$ and the back-sweep to $0\,\si{mT}$ a linear behavior is found. ANE should thus lead to a splitting of the signal at zero field. Taking into account both, AMS and ANE, leads to the grey curve in Fig.~\ref{iii}B. Clearly, this splitting predicted at low fields is not observed in the experiment. From that we can conclude that the ANE is not significant in the experimental data and seems to be overestimated by the model. Note that the temperature model is based on a wire with sharp rectangular cross sections and a sharp V-shaped notch. This leads to an overestimation of the out-of-plane gradients at the notch and hence of the ANE contribution compared to the real device with rounded edges and smooth notch (cf.\ Fig.~\ref{i}B, inset).

Similar results have been obtained on various devices with varying geometries confirming that a slight variation of the nanowire width or the notch shape does not change the general behavior. Furthermore, no significant difference between head-to head and tail-to-tail DWs was found. Our data thus clearly reveal the thermopower contribution of an individual DW in a magnetic nanowire thereby providing the fundamental link between macroscopic thermoelectrical signature and nanomagnetic spin configuration.

\bibliography{TDW_bib}

\begin{thebibliography}{38}
\expandafter\ifx\csname natexlab\endcsname\relax\def\natexlab#1{#1}\fi
\expandafter\ifx\csname bibnamefont\endcsname\relax
  \def\bibnamefont#1{#1}\fi
\expandafter\ifx\csname bibfnamefont\endcsname\relax
  \def\bibfnamefont#1{#1}\fi
\expandafter\ifx\csname citenamefont\endcsname\relax
  \def\citenamefont#1{#1}\fi
\expandafter\ifx\csname url\endcsname\relax
  \def\url#1{\texttt{#1}}\fi
\expandafter\ifx\csname urlprefix\endcsname\relax\def\urlprefix{URL }\fi
\providecommand{\bibinfo}[2]{#2}
\providecommand{\eprint}[2][]{\url{#2}}

\bibitem[{\citenamefont{Mott and Jones}(1953)}]{Mott1953book}
\bibinfo{author}{\bibfnamefont{N.~F.} \bibnamefont{Mott}} \bibnamefont{and}
  \bibinfo{author}{\bibfnamefont{H.}~\bibnamefont{Jones}},
  \emph{\bibinfo{title}{Theory of the properties of metal and alloys}}
  (\bibinfo{publisher}{Oxford University Press}, \bibinfo{year}{1953}).

\bibitem[{\citenamefont{Zutic et~al.}(2004)\citenamefont{Zutic, Fabian, and
  Sarma}}]{Zutic2004rmp}
\bibinfo{author}{\bibfnamefont{I.}~\bibnamefont{Zutic}},
  \bibinfo{author}{\bibfnamefont{J.}~\bibnamefont{Fabian}}, \bibnamefont{and}
  \bibinfo{author}{\bibfnamefont{S.~D.} \bibnamefont{Sarma}},
  \bibinfo{journal}{Rev. Mod. Phys.} \textbf{\bibinfo{volume}{76}}
  (\bibinfo{year}{2004}).

\bibitem[{\citenamefont{Johnson and Silsbee}(1987)}]{Johnson1987prb}
\bibinfo{author}{\bibfnamefont{M.}~\bibnamefont{Johnson}} \bibnamefont{and}
  \bibinfo{author}{\bibfnamefont{R.~H.} \bibnamefont{Silsbee}},
  \bibinfo{journal}{Phys. Rev. B} \textbf{\bibinfo{volume}{35}},
  \bibinfo{pages}{4959} (\bibinfo{year}{1987}).

\bibitem[{\citenamefont{Bauer et~al.}(2010{\natexlab{a}})\citenamefont{Bauer,
  MacDonald, and Maekawa}}]{Bauer2010ssc}
\bibinfo{author}{\bibfnamefont{G.~E.~W.} \bibnamefont{Bauer}},
  \bibinfo{author}{\bibfnamefont{A.~H.} \bibnamefont{MacDonald}},
  \bibnamefont{and} \bibinfo{author}{\bibfnamefont{S.}~\bibnamefont{Maekawa}},
  \bibinfo{journal}{Solid State Commun.} \textbf{\bibinfo{volume}{150}},
  \bibinfo{pages}{459} (\bibinfo{year}{2010}{\natexlab{a}}).

\bibitem[{\citenamefont{Bauer et~al.}(2012)\citenamefont{Bauer, Saitoh, and van
  Wees}}]{Bauer2012nmat}
\bibinfo{author}{\bibfnamefont{G.~E.~W.} \bibnamefont{Bauer}},
  \bibinfo{author}{\bibfnamefont{E.}~\bibnamefont{Saitoh}}, \bibnamefont{and}
  \bibinfo{author}{\bibfnamefont{B.~J.} \bibnamefont{van Wees}},
  \bibinfo{journal}{Nature Materials} \textbf{\bibinfo{volume}{11}},
  \bibinfo{pages}{391} (\bibinfo{year}{2012}).

\bibitem[{\citenamefont{Uchida}(2008)}]{Uchida2008nature}
\bibinfo{author}{\bibfnamefont{K.}~\bibnamefont{Uchida}},
  \bibinfo{journal}{Nature} \textbf{\bibinfo{volume}{455}},
  \bibinfo{pages}{778} (\bibinfo{year}{2008}).

\bibitem[{\citenamefont{Slachter et~al.}(2010)\citenamefont{Slachter, Bakker,
  Adam, and van Wees}}]{Slachter2010nphys}
\bibinfo{author}{\bibfnamefont{A.}~\bibnamefont{Slachter}},
  \bibinfo{author}{\bibfnamefont{F.~L.} \bibnamefont{Bakker}},
  \bibinfo{author}{\bibfnamefont{J.~P.} \bibnamefont{Adam}}, \bibnamefont{and}
  \bibinfo{author}{\bibfnamefont{B.~J.} \bibnamefont{van Wees}},
  \bibinfo{journal}{Nature Phys.} \textbf{\bibinfo{volume}{6}},
  \bibinfo{pages}{879} (\bibinfo{year}{2010}).

\bibitem[{\citenamefont{LeBreton et~al.}(2011)\citenamefont{LeBreton, Sharma,
  Saito, Yuasa, and Jansen}}]{LeBreton2011nature}
\bibinfo{author}{\bibfnamefont{J.-C.} \bibnamefont{LeBreton}},
  \bibinfo{author}{\bibfnamefont{S.}~\bibnamefont{Sharma}},
  \bibinfo{author}{\bibfnamefont{H.}~\bibnamefont{Saito}},
  \bibinfo{author}{\bibfnamefont{S.}~\bibnamefont{Yuasa}}, \bibnamefont{and}
  \bibinfo{author}{\bibfnamefont{R.}~\bibnamefont{Jansen}},
  \bibinfo{journal}{Nature} \textbf{\bibinfo{volume}{475}}, \bibinfo{pages}{82}
  (\bibinfo{year}{2011}).

\bibitem[{\citenamefont{Jeon et~al.}(2014)\citenamefont{Jeon, Min, Spiesser,
  Saito, Shin, Yuasa, and Jansen}}]{Jeon2014nmat}
\bibinfo{author}{\bibfnamefont{K.-R.} \bibnamefont{Jeon}},
  \bibinfo{author}{\bibfnamefont{B.-C.} \bibnamefont{Min}},
  \bibinfo{author}{\bibfnamefont{A.}~\bibnamefont{Spiesser}},
  \bibinfo{author}{\bibfnamefont{H.}~\bibnamefont{Saito}},
  \bibinfo{author}{\bibfnamefont{S.-C.} \bibnamefont{Shin}},
  \bibinfo{author}{\bibfnamefont{S.}~\bibnamefont{Yuasa}}, \bibnamefont{and}
  \bibinfo{author}{\bibfnamefont{R.}~\bibnamefont{Jansen}},
  \bibinfo{journal}{Nature Materials} \textbf{\bibinfo{volume}{13}},
  \bibinfo{pages}{360} (\bibinfo{year}{2014}).

\bibitem[{\citenamefont{Avery et~al.}(2012)\citenamefont{Avery, Pufall, and
  Zink}}]{Avery2012prl}
\bibinfo{author}{\bibfnamefont{A.~D.} \bibnamefont{Avery}},
  \bibinfo{author}{\bibfnamefont{M.~R.} \bibnamefont{Pufall}},
  \bibnamefont{and} \bibinfo{author}{\bibfnamefont{B.~L.} \bibnamefont{Zink}},
  \bibinfo{journal}{Phys. Rev. Lett.} \textbf{\bibinfo{volume}{109}},
  \bibinfo{pages}{196602} (\bibinfo{year}{2012}).

\bibitem[{\citenamefont{Schmid et~al.}(2013)\citenamefont{Schmid, Srichandan,
  Meier, Kuschel, Schmalhorst, Vogel, Reiss, Strunk, and Back}}]{Schmid2013prl}
\bibinfo{author}{\bibfnamefont{M.}~\bibnamefont{Schmid}},
  \bibinfo{author}{\bibfnamefont{S.}~\bibnamefont{Srichandan}},
  \bibinfo{author}{\bibfnamefont{D.}~\bibnamefont{Meier}},
  \bibinfo{author}{\bibfnamefont{T.}~\bibnamefont{Kuschel}},
  \bibinfo{author}{\bibfnamefont{J.-M.} \bibnamefont{Schmalhorst}},
  \bibinfo{author}{\bibfnamefont{M.}~\bibnamefont{Vogel}},
  \bibinfo{author}{\bibfnamefont{G.}~\bibnamefont{Reiss}},
  \bibinfo{author}{\bibfnamefont{C.}~\bibnamefont{Strunk}}, \bibnamefont{and}
  \bibinfo{author}{\bibfnamefont{C.~H.} \bibnamefont{Back}},
  \bibinfo{journal}{Phys. Rev. Lett.} \textbf{\bibinfo{volume}{111}},
  \bibinfo{pages}{187201} (\bibinfo{year}{2013}).

\bibitem[{\citenamefont{Shi et~al.}(1993)\citenamefont{Shi, Yu, Parkin, and
  Salamon}}]{Shi1993jap}
\bibinfo{author}{\bibfnamefont{J.}~\bibnamefont{Shi}},
  \bibinfo{author}{\bibfnamefont{R.~C.} \bibnamefont{Yu}},
  \bibinfo{author}{\bibfnamefont{S.~S.~P.} \bibnamefont{Parkin}},
  \bibnamefont{and} \bibinfo{author}{\bibfnamefont{M.~B.}
  \bibnamefont{Salamon}}, \bibinfo{journal}{J. Appl. Phys.}
  \textbf{\bibinfo{volume}{73}}, \bibinfo{pages}{5524} (\bibinfo{year}{1993}).

\bibitem[{\citenamefont{Czerner et~al.}(2011)\citenamefont{Czerner, Bachmann,
  and Heiliger}}]{Czerner2011prb}
\bibinfo{author}{\bibfnamefont{M.}~\bibnamefont{Czerner}},
  \bibinfo{author}{\bibfnamefont{M.}~\bibnamefont{Bachmann}}, \bibnamefont{and}
  \bibinfo{author}{\bibfnamefont{C.}~\bibnamefont{Heiliger}},
  \bibinfo{journal}{Phys. Rev. B} \textbf{\bibinfo{volume}{83}},
  \bibinfo{pages}{132405} (\bibinfo{year}{2011}).

\bibitem[{\citenamefont{Walter et~al.}(2011)\citenamefont{Walter, Walowski,
  Zbarsky, Muenzenberg, Schaefers, Ebke, Reiss, Thomas, Peretzki, Seibt
  et~al.}}]{Walter2011nmat}
\bibinfo{author}{\bibfnamefont{M.}~\bibnamefont{Walter}},
  \bibinfo{author}{\bibfnamefont{J.}~\bibnamefont{Walowski}},
  \bibinfo{author}{\bibfnamefont{V.}~\bibnamefont{Zbarsky}},
  \bibinfo{author}{\bibfnamefont{M.}~\bibnamefont{Muenzenberg}},
  \bibinfo{author}{\bibfnamefont{M.}~\bibnamefont{Schaefers}},
  \bibinfo{author}{\bibfnamefont{D.}~\bibnamefont{Ebke}},
  \bibinfo{author}{\bibfnamefont{G.}~\bibnamefont{Reiss}},
  \bibinfo{author}{\bibfnamefont{A.}~\bibnamefont{Thomas}},
  \bibinfo{author}{\bibfnamefont{P.}~\bibnamefont{Peretzki}},
  \bibinfo{author}{\bibfnamefont{M.}~\bibnamefont{Seibt}},
  \bibnamefont{et~al.}, \bibinfo{journal}{Nature Materials}
  \textbf{\bibinfo{volume}{10}}, \bibinfo{pages}{742} (\bibinfo{year}{2011}).

\bibitem[{\citenamefont{Liebing et~al.}(2011)\citenamefont{Liebing,
  Serrano-Guisan, Rott, Reiss, Langer, Ocker, and Schumacher}}]{Liebing2011prl}
\bibinfo{author}{\bibfnamefont{N.}~\bibnamefont{Liebing}},
  \bibinfo{author}{\bibfnamefont{S.}~\bibnamefont{Serrano-Guisan}},
  \bibinfo{author}{\bibfnamefont{K.}~\bibnamefont{Rott}},
  \bibinfo{author}{\bibfnamefont{G.}~\bibnamefont{Reiss}},
  \bibinfo{author}{\bibfnamefont{J.}~\bibnamefont{Langer}},
  \bibinfo{author}{\bibfnamefont{B.}~\bibnamefont{Ocker}}, \bibnamefont{and}
  \bibinfo{author}{\bibfnamefont{H.~W.} \bibnamefont{Schumacher}},
  \bibinfo{journal}{Phys. Rev. Lett.} \textbf{\bibinfo{volume}{107}},
  \bibinfo{pages}{177201} (\bibinfo{year}{2011}).

\bibitem[{\citenamefont{B{\"o}hnert et~al.}(2013)\citenamefont{B{\"o}hnert,
  Vega, Michel, Prida, and Nielsch}}]{Bohnert2013apl}
\bibinfo{author}{\bibfnamefont{T.}~\bibnamefont{B{\"o}hnert}},
  \bibinfo{author}{\bibfnamefont{V.}~\bibnamefont{Vega}},
  \bibinfo{author}{\bibfnamefont{A.-K.} \bibnamefont{Michel}},
  \bibinfo{author}{\bibfnamefont{V.~M.} \bibnamefont{Prida}}, \bibnamefont{and}
  \bibinfo{author}{\bibfnamefont{K.}~\bibnamefont{Nielsch}},
  \bibinfo{journal}{Appl. Phys. Lett.} \textbf{\bibinfo{volume}{103}},
  \bibinfo{pages}{092407} (\bibinfo{year}{2013}).

\bibitem[{\citenamefont{Gravier et~al.}(2006)\citenamefont{Gravier,
  Serrano-Guisan, Reuse, and Ansermet}}]{Gravier2006prb}
\bibinfo{author}{\bibfnamefont{L.}~\bibnamefont{Gravier}},
  \bibinfo{author}{\bibfnamefont{S.}~\bibnamefont{Serrano-Guisan}},
  \bibinfo{author}{\bibfnamefont{F.}~\bibnamefont{Reuse}}, \bibnamefont{and}
  \bibinfo{author}{\bibfnamefont{J.~P.} \bibnamefont{Ansermet}},
  \bibinfo{journal}{Phys. Rev. B} \textbf{\bibinfo{volume}{73}},
  \bibinfo{pages}{024419} (\bibinfo{year}{2006}).

\bibitem[{\citenamefont{Ono et~al.}(1999)\citenamefont{Ono, Miyajima, Shigeto,
  Mibu, Hosoito, and Shinjo}}]{Ono1999science}
\bibinfo{author}{\bibfnamefont{T.}~\bibnamefont{Ono}},
  \bibinfo{author}{\bibfnamefont{H.}~\bibnamefont{Miyajima}},
  \bibinfo{author}{\bibfnamefont{K.}~\bibnamefont{Shigeto}},
  \bibinfo{author}{\bibfnamefont{K.}~\bibnamefont{Mibu}},
  \bibinfo{author}{\bibfnamefont{N.}~\bibnamefont{Hosoito}}, \bibnamefont{and}
  \bibinfo{author}{\bibfnamefont{T.}~\bibnamefont{Shinjo}},
  \bibinfo{journal}{Science} \textbf{\bibinfo{volume}{284}},
  \bibinfo{pages}{486} (\bibinfo{year}{1999}).

\bibitem[{\citenamefont{Allwood et~al.}(2005)\citenamefont{Allwood, Xiong,
  Faulkner, Atkinson, Petit, and Cowburn}}]{Allwood2005science}
\bibinfo{author}{\bibfnamefont{D.~A.} \bibnamefont{Allwood}},
  \bibinfo{author}{\bibfnamefont{G.}~\bibnamefont{Xiong}},
  \bibinfo{author}{\bibfnamefont{C.~C.} \bibnamefont{Faulkner}},
  \bibinfo{author}{\bibfnamefont{D.}~\bibnamefont{Atkinson}},
  \bibinfo{author}{\bibfnamefont{D.}~\bibnamefont{Petit}}, \bibnamefont{and}
  \bibinfo{author}{\bibfnamefont{R.~P.} \bibnamefont{Cowburn}},
  \bibinfo{journal}{Science} \textbf{\bibinfo{volume}{309}},
  \bibinfo{pages}{1688} (\bibinfo{year}{2005}).

\bibitem[{\citenamefont{Parkin et~al.}(2008)\citenamefont{Parkin, Hayashi, and
  Thomas}}]{Parkin2008science}
\bibinfo{author}{\bibfnamefont{S.~S.~P.} \bibnamefont{Parkin}},
  \bibinfo{author}{\bibfnamefont{M.}~\bibnamefont{Hayashi}}, \bibnamefont{and}
  \bibinfo{author}{\bibfnamefont{L.}~\bibnamefont{Thomas}},
  \bibinfo{journal}{Science} \textbf{\bibinfo{volume}{320}},
  \bibinfo{pages}{190} (\bibinfo{year}{2008}).

\bibitem[{\citenamefont{Berger}(1985)}]{Berger1985apl}
\bibinfo{author}{\bibfnamefont{L.}~\bibnamefont{Berger}}, \bibinfo{journal}{J.
  Appl. Phys.} \textbf{\bibinfo{volume}{58}}, \bibinfo{pages}{450}
  (\bibinfo{year}{1985}).

\bibitem[{\citenamefont{Hatami et~al.}(2007)\citenamefont{Hatami, Bauer, Zhang,
  and Kelly}}]{Hatami2007prl}
\bibinfo{author}{\bibfnamefont{M.}~\bibnamefont{Hatami}},
  \bibinfo{author}{\bibfnamefont{G.~E.~W.} \bibnamefont{Bauer}},
  \bibinfo{author}{\bibfnamefont{Q.}~\bibnamefont{Zhang}}, \bibnamefont{and}
  \bibinfo{author}{\bibfnamefont{P.~J.} \bibnamefont{Kelly}},
  \bibinfo{journal}{Phys. Rev. Lett.} \textbf{\bibinfo{volume}{99}},
  \bibinfo{pages}{066603} (\bibinfo{year}{2007}).

\bibitem[{\citenamefont{Kovalev and Tserkovnyak}(2009)}]{Kovalev2009prb}
\bibinfo{author}{\bibfnamefont{A.}~\bibnamefont{Kovalev}} \bibnamefont{and}
  \bibinfo{author}{\bibfnamefont{Y.}~\bibnamefont{Tserkovnyak}},
  \bibinfo{journal}{Phys. Rev. B} \textbf{\bibinfo{volume}{80}},
  \bibinfo{pages}{100408R} (\bibinfo{year}{2009}).

\bibitem[{\citenamefont{Hinzke and Nowak}(2011)}]{Hinzke2011prl}
\bibinfo{author}{\bibfnamefont{D.}~\bibnamefont{Hinzke}} \bibnamefont{and}
  \bibinfo{author}{\bibfnamefont{U.}~\bibnamefont{Nowak}},
  \bibinfo{journal}{Phys. Rev. Lett.} \textbf{\bibinfo{volume}{107}},
  \bibinfo{pages}{027205} (\bibinfo{year}{2011}).

\bibitem[{\citenamefont{Yan et~al.}(2011)\citenamefont{Yan, Wang, and
  Wang}}]{Yan2011prl}
\bibinfo{author}{\bibfnamefont{P.}~\bibnamefont{Yan}},
  \bibinfo{author}{\bibfnamefont{X.~S.} \bibnamefont{Wang}}, \bibnamefont{and}
  \bibinfo{author}{\bibfnamefont{X.~R.} \bibnamefont{Wang}},
  \bibinfo{journal}{Phys. Rev. Lett.} \textbf{\bibinfo{volume}{107}},
  \bibinfo{pages}{177207} (\bibinfo{year}{2011}).

\bibitem[{\citenamefont{Torrejon et~al.}(2013)\citenamefont{Torrejon,
  Malinowski, Pelloux, Weil, Thiaville, Curiale, Lacour, Montaigne, and
  Hehn}}]{Torrejon2013prl}
\bibinfo{author}{\bibfnamefont{J.}~\bibnamefont{Torrejon}},
  \bibinfo{author}{\bibfnamefont{G.}~\bibnamefont{Malinowski}},
  \bibinfo{author}{\bibfnamefont{M.}~\bibnamefont{Pelloux}},
  \bibinfo{author}{\bibfnamefont{R.}~\bibnamefont{Weil}},
  \bibinfo{author}{\bibfnamefont{A.}~\bibnamefont{Thiaville}},
  \bibinfo{author}{\bibfnamefont{J.}~\bibnamefont{Curiale}},
  \bibinfo{author}{\bibfnamefont{D.}~\bibnamefont{Lacour}},
  \bibinfo{author}{\bibfnamefont{F.}~\bibnamefont{Montaigne}},
  \bibnamefont{and} \bibinfo{author}{\bibfnamefont{M.}~\bibnamefont{Hehn}},
  \bibinfo{journal}{Phys. Rev. Lett.} \textbf{\bibinfo{volume}{110}},
  \bibinfo{pages}{177202} (\bibinfo{year}{2013}).

\bibitem[{\citenamefont{Jiang et~al.}(2013)\citenamefont{Jiang, Upadhyaya, Fan,
  Zhao, Wang, Chang, Lang, Wong, Lewis, Lin et~al.}}]{Jiang2013prl}
\bibinfo{author}{\bibfnamefont{W.}~\bibnamefont{Jiang}},
  \bibinfo{author}{\bibfnamefont{P.}~\bibnamefont{Upadhyaya}},
  \bibinfo{author}{\bibfnamefont{Y.}~\bibnamefont{Fan}},
  \bibinfo{author}{\bibfnamefont{J.}~\bibnamefont{Zhao}},
  \bibinfo{author}{\bibfnamefont{M.}~\bibnamefont{Wang}},
  \bibinfo{author}{\bibfnamefont{L.-T.} \bibnamefont{Chang}},
  \bibinfo{author}{\bibfnamefont{M.}~\bibnamefont{Lang}},
  \bibinfo{author}{\bibfnamefont{K.~L.} \bibnamefont{Wong}},
  \bibinfo{author}{\bibfnamefont{M.}~\bibnamefont{Lewis}},
  \bibinfo{author}{\bibfnamefont{Y.-T.} \bibnamefont{Lin}},
  \bibnamefont{et~al.}, \bibinfo{journal}{Phys. Rev. Lett.}
  \textbf{\bibinfo{volume}{110}}, \bibinfo{pages}{177202}
  (\bibinfo{year}{2013}).

\bibitem[{\citenamefont{Bauer et~al.}(2010{\natexlab{b}})\citenamefont{Bauer,
  Bretzel, Brataas, and Tserkovnyak}}]{Bauer2010prb}
\bibinfo{author}{\bibfnamefont{G.~E.~W.} \bibnamefont{Bauer}},
  \bibinfo{author}{\bibfnamefont{S.}~\bibnamefont{Bretzel}},
  \bibinfo{author}{\bibfnamefont{A.}~\bibnamefont{Brataas}}, \bibnamefont{and}
  \bibinfo{author}{\bibfnamefont{Y.}~\bibnamefont{Tserkovnyak}},
  \bibinfo{journal}{Phys. Rev. B} \textbf{\bibinfo{volume}{81}},
  \bibinfo{pages}{024427} (\bibinfo{year}{2010}{\natexlab{b}}).

\bibitem[{\citenamefont{Corte-Leon et~al.}(2014)\citenamefont{Corte-Leon,
  Nabaei, Manzin, Fletcher, Krzysteczko, Schumacher, and
  Kazakova}}]{Corte2014srep}
\bibinfo{author}{\bibfnamefont{H.}~\bibnamefont{Corte-Leon}},
  \bibinfo{author}{\bibfnamefont{V.}~\bibnamefont{Nabaei}},
  \bibinfo{author}{\bibfnamefont{A.}~\bibnamefont{Manzin}},
  \bibinfo{author}{\bibfnamefont{J.}~\bibnamefont{Fletcher}},
  \bibinfo{author}{\bibfnamefont{P.}~\bibnamefont{Krzysteczko}},
  \bibinfo{author}{\bibfnamefont{H.~W.} \bibnamefont{Schumacher}},
  \bibnamefont{and} \bibinfo{author}{\bibfnamefont{O.}~\bibnamefont{Kazakova}},
  \bibinfo{journal}{Sci. Rep.} \textbf{\bibinfo{volume}{4}},
  \bibinfo{pages}{6045} (\bibinfo{year}{2014}).

\bibitem[{\citenamefont{Scheinfein et~al.}(1991)\citenamefont{Scheinfein,
  Unguris, Blue, Coakley, Pierce, Celotta, and Ryan}}]{Scheinfein1991prb}
\bibinfo{author}{\bibfnamefont{M.~R.} \bibnamefont{Scheinfein}},
  \bibinfo{author}{\bibfnamefont{J.}~\bibnamefont{Unguris}},
  \bibinfo{author}{\bibfnamefont{J.~L.} \bibnamefont{Blue}},
  \bibinfo{author}{\bibfnamefont{K.~J.} \bibnamefont{Coakley}},
  \bibinfo{author}{\bibfnamefont{D.}~\bibnamefont{Pierce}},
  \bibinfo{author}{\bibfnamefont{R.~J.} \bibnamefont{Celotta}},
  \bibnamefont{and} \bibinfo{author}{\bibfnamefont{P.}~\bibnamefont{Ryan}},
  \bibinfo{journal}{Phys. Rev. B} \textbf{\bibinfo{volume}{43}},
  \bibinfo{pages}{3395} (\bibinfo{year}{1991}).

\bibitem[{\citenamefont{Zhang et~al.}(2005)\citenamefont{Zhang, Xie, Fujii,
  Ago, Takahashi, Ikuta, Abe, and Shimizu}}]{Zhang2005apl}
\bibinfo{author}{\bibfnamefont{X.}~\bibnamefont{Zhang}},
  \bibinfo{author}{\bibfnamefont{H.}~\bibnamefont{Xie}},
  \bibinfo{author}{\bibfnamefont{M.}~\bibnamefont{Fujii}},
  \bibinfo{author}{\bibfnamefont{H.}~\bibnamefont{Ago}},
  \bibinfo{author}{\bibfnamefont{K.}~\bibnamefont{Takahashi}},
  \bibinfo{author}{\bibfnamefont{T.}~\bibnamefont{Ikuta}},
  \bibinfo{author}{\bibfnamefont{H.}~\bibnamefont{Abe}}, \bibnamefont{and}
  \bibinfo{author}{\bibfnamefont{T.}~\bibnamefont{Shimizu}},
  \bibinfo{journal}{Appl. Phys. Lett.} \textbf{\bibinfo{volume}{86}},
  \bibinfo{pages}{171912} (\bibinfo{year}{2005}).

\bibitem[{\citenamefont{Slachter et~al.}(2011)\citenamefont{Slachter, Bakker,
  and van Wees}}]{Slachter2011prb}
\bibinfo{author}{\bibfnamefont{A.}~\bibnamefont{Slachter}},
  \bibinfo{author}{\bibfnamefont{F.~L.} \bibnamefont{Bakker}},
  \bibnamefont{and} \bibinfo{author}{\bibfnamefont{B.~J.} \bibnamefont{van
  Wees}}, \bibinfo{journal}{Phys. Rev. B} \textbf{\bibinfo{volume}{84}},
  \bibinfo{pages}{020412} (\bibinfo{year}{2011}).

\bibitem[{\citenamefont{Soni and Okram}(2008)}]{Soni2008rsi}
\bibinfo{author}{\bibfnamefont{A.}~\bibnamefont{Soni}} \bibnamefont{and}
  \bibinfo{author}{\bibfnamefont{G.~S.} \bibnamefont{Okram}},
  \bibinfo{journal}{Rev. Sci. Inst} \textbf{\bibinfo{volume}{79}},
  \bibinfo{pages}{125103} (\bibinfo{year}{2008}).

\bibitem[{\citenamefont{Ho et~al.}(1978)\citenamefont{Ho, Ackerman, Wu, Oh, and
  Havill}}]{Ho1978jpc}
\bibinfo{author}{\bibfnamefont{C.~Y.} \bibnamefont{Ho}},
  \bibinfo{author}{\bibfnamefont{M.~W.} \bibnamefont{Ackerman}},
  \bibinfo{author}{\bibfnamefont{K.~Y.} \bibnamefont{Wu}},
  \bibinfo{author}{\bibfnamefont{S.~G.} \bibnamefont{Oh}}, \bibnamefont{and}
  \bibinfo{author}{\bibfnamefont{T.~N.} \bibnamefont{Havill}},
  \bibinfo{journal}{J. Phys. Chem. Ref. Data} \textbf{\bibinfo{volume}{7}},
  \bibinfo{pages}{959} (\bibinfo{year}{1978}).

\bibitem[{\citenamefont{Owen et~al.}(1937)\citenamefont{Owen, Yates, and
  Sully}}]{Owen1937pps}
\bibinfo{author}{\bibfnamefont{E.~A.} \bibnamefont{Owen}},
  \bibinfo{author}{\bibfnamefont{E.~L.} \bibnamefont{Yates}}, \bibnamefont{and}
  \bibinfo{author}{\bibfnamefont{A.~H.} \bibnamefont{Sully}},
  \bibinfo{journal}{Proc. Phys. Soc.} \textbf{\bibinfo{volume}{49}},
  \bibinfo{pages}{315} (\bibinfo{year}{1937}).

\bibitem[{\citenamefont{Bonnenberg et~al.}(2000)\citenamefont{Bonnenberg,
  Hempel, and Wijn}}]{Bonnenberg2000book}
\bibinfo{author}{\bibfnamefont{D.}~\bibnamefont{Bonnenberg}},
  \bibinfo{author}{\bibfnamefont{K.~A.} \bibnamefont{Hempel}},
  \bibnamefont{and} \bibinfo{author}{\bibfnamefont{H.~P.~J.}
  \bibnamefont{Wijn}}, \emph{\bibinfo{title}{Springer Materials: The
  Landolt-Br\"onstein Database}} (\bibinfo{publisher}{Springer},
  \bibinfo{year}{2000}).

\bibitem[{\citenamefont{Hankenmeier et~al.}(2008)\citenamefont{Hankenmeier,
  Sachse, Stark, Fr\"omter, and Oepen}}]{Hankenmeier2008apl}
\bibinfo{author}{\bibfnamefont{S.}~\bibnamefont{Hankenmeier}},
  \bibinfo{author}{\bibfnamefont{K.}~\bibnamefont{Sachse}},
  \bibinfo{author}{\bibfnamefont{Y.}~\bibnamefont{Stark}},
  \bibinfo{author}{\bibfnamefont{R.}~\bibnamefont{Fr\"omter}},
  \bibnamefont{and} \bibinfo{author}{\bibfnamefont{H.}~\bibnamefont{Oepen}},
  \bibinfo{journal}{Appl. Phys. Lett.} \textbf{\bibinfo{volume}{92}},
  \bibinfo{pages}{242503} (\bibinfo{year}{2008}).

\bibitem[{\citenamefont{Banhart and Ebert}(1995)}]{Banhart1995epl}
\bibinfo{author}{\bibfnamefont{J.}~\bibnamefont{Banhart}} \bibnamefont{and}
  \bibinfo{author}{\bibfnamefont{H.}~\bibnamefont{Ebert}},
  \bibinfo{journal}{Europhys. Lett.} \textbf{\bibinfo{volume}{32}},
  \bibinfo{pages}{517} (\bibinfo{year}{1995}).

\end{thebibliography}

\newpage

\section*{Supplementary material}

\subsection{Device Fabrication}

In our experiments we use an L-shaped Py nanowire with a notch. The nanowire is $290\,\si{nm}$ wide and has arms of $2\,\si{\micro m}$ and $4\,\si{\micro m}$ length. The longer arm has a notch, $150\,\si{nm}$ deep and $280\,\si{nm}$ wide, at a distance of $3\,\si{\micro m}$ from the corner. The nanostructure is patterned by electron beam lithography in combination with Ar ion etching from a continuous Py film that has been sputter-deposited on a $525\,\si{\micro m}$ Si substrate covered by a $50\,\si{nm}$ SiO layer. The Py is $27\,\si{nm}$ thick and covered with a Pt cap of $2\,\si{nm}$ to prevent oxidation. Additionally, devices without Pt cap were fabricated to ensure that the Pt capping layer has no significant influence on the DWTP. In a second lithography step, we attach Pt wires as voltage probes. The Pt wires are $115\,\si{nm}$ thick with a $10\,\si{nm}$ Ta adhesion layer. The interface between Py and Ta is cleaned in-situ by low energy Ar ions prior to Ta/Pt deposition to ensure good electrical contact. Two additional Pt strips located at a distance of $0.5\,\si{\micro m}$ and $1.5\,\si{\micro m}$ from the Py nanowire serve as resistive thermometer and heater, respectively.

\subsection{Temperature calibration}

To detect the temperature gradient experimentally, at least two thermometers are needed, each sensitive to the temperature $T_d$ at a certain distance $d$ from the heat source. We fabricate a set of nominally identical devices with heater-thermometer pairs separated by \SIlist{1;2;4}{\micro m}. For each distance, the calibration is repeated for four different devices to increase the statistical significance. The temperature coefficient of the Pt thermometer $\alpha_\mathrm{Pt}$ is measured by 4-wire resistance $R$ measurements as a function of the temperature $T$ of a hot plate heated to up to $30\,\si{K}$ above room temperature. We use thermal grease and an equilibration time of at least $30\,\si{min}$ to ensure uniform temperature distribution before taking a $R(T)$ value. The resulting temperature coefficient $0.00125\,\si{K^{-1}}<\alpha_\mathrm{Pt}<0.00135\,\si{K^{-1}}$ allows for a measurement of the local temperature increase $\delta T$ with an accuracy of approx.\ $\pm4\,\%$. This results in an uncertainty of the temperature difference $\Delta T$ measured between $d=4\,\si{\micro m}$ and $d=2\,\si{\micro m}$ of approx.\ $\pm20\,\%$. This uncertainty is not to be confused with poor time-stability.

\subsection{The influence of the wiring}

The 2-wire AMR ratio $(R_\parallel-R_\perp)/R_\perp$ is $0.45\,\%$. This value, however, is obscured by the resistance of the wiring which is estimated on the basis of 4-wire measurements on similar devices to $R_\mathrm{wire}=195.6\,\si{\ohm}$. This yields a 4-wire AMR ratio of $1.4\,\%$.

Also the measured Seebeck coefficients are influenced by the electrical contacts. Considering a temperature gradient with $\Delta T=2.4\,\si{K}$, the nominal Seebeck coefficient of the permalloy-platinum thermocouple is $S_\parallel=-V_\parallel/\Delta T=-23.4\,\si{\micro V/K}$ and $S_\perp=-V_\perp/\Delta T=-23.6\,\si{\micro V/K}$ for the longitudinal and transversal geometry, respectively. This yields a magneto-Seebeck ratio $(S_\parallel-S_\perp)/S_\perp$ of $-0.8\,\%$. The Pt voltage probe contributes its own thermopower of $10.8\,\si{\micro V}$ due to a Seebeck coefficient of $S_\mathrm{Pt}=-4.5\,\si{\micro V/K}$ \cite{Soni2008rsi}. The resulting absolute Seebeck coefficients of permalloy are $S_\parallel^\mathrm{Py}=-18.9\,\si{\micro V/K}$ and $S_\perp^\mathrm{Py}=-19.1\,\si{\micro V/K}$ with an uncertainty of $\pm5\,\si{\micro V/K}$. This yields an absolute magneto-Seebeck ratio of $(-1.0\pm0.05)\,\%$.

\subsection{Modeling}

For micromagnetic simulations we use a commercial micromagnetic modeling tool (\textsc{llg} Micromagnetics Simulator) \cite{Scheinfein1991prb} with the following parameters: saturation magnetization $\mu_0M_S=1.005\,\si{T}$, exchange stiffness $A=1.05\times10^{-11}\,\si{J/m}$, and uniaxial anisotropy constant $K_u=100\,\si{J\per\cubic\metre}$ oriented along the long wire. The cells size is $10\times10\times25\,\si{\cubic\nm}$. The simulation temperature is zero Kelvin.

\begin{table}
\begin{tabular}{llll}
\hline
\hline
  Parameter & Permalloy & SiO & Si\\
  \hline
  Thermal conductivity [$\si{W/K\,m}$]\hspace{1cm}	& 46.4 \cite{Ho1978jpc}\hspace{1.5cm}		& 1.4\hspace{1.5cm}	& 130\hspace{1.5cm}\\
  Density [$\si{kg/m^{3}}$] 							& 8700 \cite{Owen1937pps} 				& 2200				& 2329\\
  Heat capacity [$\si{J/kg\,K}$]	 					& 430 \cite{Bonnenberg2000book} 			& 730				& 700\\
  Electrical conductivity [$\si{S/m}$] 					& $4\times10^6$ \cite{Hankenmeier2008apl} 	& 0					& $4\times 10^{-12}$\\
\hline
\hline
\end{tabular}
\caption{Material parameters used for calculation of temperature distribution. Values for Si and SiO are taken directly from the \textsc{comsol} software. For Py we use literature values, as indicated.}
\label{tab1}
\end{table}

The temperature distribution is modeled by a commercial finite-element modeling tool (\textsc{comsol} Multiphysics). The input parameters are displayed in Tab.~\ref{tab1}. The boundary condition `convective cooling' is activated for all surfaces in contact with air. The temperature of the air and of the bottom surface of the SiO substrate is fixed at $300\,\si{K}$. 

\subsection{The Seebeck tensor}

An n-type conductor placed in a temperature gradient $\nabla T$ generally accumulates negative charge at the cool side leading to an electrical field $\vec{E}$ pointing away from the heat source. The efficiency of this process is described by the Seebeck coefficient $S$ with $\vec{E}=S\,\nabla T$ (by this definition the Seebeck coefficient is negative for n-type conductors). It is customary to use $\vec{E}=-\nabla T_\mathrm{T}$ and rewrite this equation to $\nabla V_\mathrm{T}=-S\,\nabla T$ which by integration directly leads to the measured voltage $V_\mathrm{T}=-S\,\Delta T$. This voltage, which is referred to as thermopower, is generated between two points with a temperature difference of $\Delta T$.

Our phenomenological description of the magneto-Seebeck effect is based on the according description of the magnetoresistance \cite{Banhart1995epl}. The Seebeck tensor for systems with the magnetization along the $x$-direction has the form
\begin{equation}
\nabla V_\mathrm{T}=-
\begin{pmatrix}
S_\parallel &  & \\
 & S_\perp & -S_N\\
 & S_N & S_\perp
\end{pmatrix}
\nabla T\,
\label{eq_matrix}
\end{equation}
where $S_N$ is a measure of the anomalous Nernst effect. $S_\parallel$ and $S_\perp$ are the longitudinal and transversal Seebeck coefficients, respectively. For a magnetization vector pointing in an arbitrary direction in the $xy$-plane we need to transform the Seebeck tensor $\mathsf{S}$ using the rotational matrix $\mathsf{D}_\theta$ with $\theta$ the angle of the magnetization in respect to the $x$-axis
\begin{equation}
\nabla V_\mathrm{T}=-\mathsf{D}_\theta\mathsf{S}\mathsf{D}^{-1}_\theta\,\nabla T\,. 
\end{equation}
Using $\Delta S=S_\parallel-S_\perp$ this result can be written as
\begin{equation}
\nabla V_\mathrm{T}=-
\begin{pmatrix}
S_\perp+\Delta S\cos^2(\theta) \\
\Delta S\cos(\theta)\sin(\theta) \\
-S_N\sin(\theta)
\end{pmatrix}
\partial_xT-
\begin{pmatrix}
\Delta S\cos(\theta)\sin(\theta) \\
S_\perp+\Delta S\sin^2(\theta) \\
S_N\cos(\theta)
\end{pmatrix}
\partial_yT-
\begin{pmatrix}
S_N\sin(\theta) \\
-S_N\cos(\theta) \\
S_\perp
\end{pmatrix}
\partial_zT\,.
\end{equation}
In our devices we are sensitive to the $x$-component of this vector which yields
\begin{equation}
 V_\mathrm{T} = -\left(S_\perp+\Delta S\cos^2(\theta)\right)\Delta T_x -\Delta S\cos(\theta)\sin(\theta)\,\Delta T_y -S_N\sin(\theta)\,\Delta T_z\,.                      
\end{equation}
We use the experimental results $S_\perp=-23.6\,\si{\micro V/K}$ and $\Delta S=0.2\,\si{\micro V/K}$ as well as $S_N=-2.6\,\si{\micro V\per K}$ taken from literature \cite{Slachter2011prb}.

\end{document}